\documentclass[amsmath,amssymb,twocolumn,aps,pre]{revtex4-1}

\usepackage{default}
\usepackage{amssymb}
\usepackage{color}
\usepackage{amsmath}
\usepackage{mathrsfs}
\usepackage{graphicx}
\usepackage{subfig}
\usepackage{epsfig}

\begin{document}

\title{Origin and stability of dark pulse Kerr combs in normal dispersion resonators}

\author{Pedro Parra-Rivas$^{1,2}$, Dami\`a Gomila$^2$, Edgar Knobloch$^3$, St\'ephane Coen$^4$ and Lendert Gelens$^{1,5,6}$}
\affiliation{
$^{1}$Applied Physics Research Group, Vrije Universiteit Brussel, 1050 Brussels, Belgium \\
$^{2}$IFISC institute (CSIC-UIB), Campus Universitat de les Illes Balears, E-07122 Palma de Mallorca, Spain \\
$^{3}$ Department of Physics, University of California, Berkeley CA 94720, USA \\
$^{4}$ Dodd-Walls Centre, and Physics Department, The University of Auckland, Private Bag 92019, Auckland~1142, New Zealand \\
$^{5}$ Laboratory of Dynamics in Biological Systems, KU Leuven, Department of Cellular and Molecular Medicine, University of Leuven, B-3000 Leuven, Belgium \\
$^{6}$ Department of Chemical and Systems Biology, Stanford University School of Medicine, Stanford CA 94305-5174, USA }

\date{\today}

\email{lendert.gelens@kuleuven.be}

\begin{abstract}
  We analyze dark pulse Kerr frequency combs in optical resonators with normal group-velocity
  dispersion using the Lugiato-Lefever model. We show that in the time domain these correspond to interlocked
  switching waves between the upper and lower homogeneous states, and explain how this fact accounts for many of their experimentally observed
  properties. Modulational instability does not play any role in their existence. Furthermore, we provide a detailed
  map indicating where stable dark pulse Kerr combs can be found in parameter space, and how they are destabilized
  for increasing values of frequency detuning.
\end{abstract}

\maketitle

\noindent Optical frequency combs generated in passive Kerr microresonators have attracted a lot of interest in
recent years for their potential for on-chip integration of frequency comb technology \cite{delhaye_optical_2007,
kippenberg_microresonator-based_2011, okawachi_octave-spanning_2011, ferdous_spectral_2011,
herr_universal_2012,papp_microresonator_2014}. Their applications span arbitrary waveform synthesis \cite{ferdous_spectral_2011},
telecommunications \cite{pfeifle_coherent_2014}, and ultra-accurate clocks \cite{papp_microresonator_2014}. These
so-called ``Kerr frequency combs'' (KFCs) are obtained by driving a high-$Q$ microresonator with a continuous-wave
(cw) laser. This leads to the generation of spectral sidebands through modulational instability (MI) and subsequent
cascaded four-wave mixing. Interestingly, much can been learned about KFCs by considering a time domain
representation. In fact, most reported KFCs correspond either to extended periodic patterns or to ultrashort pulses
known as temporal cavity solitons (CSs), stable or fluctuating \cite{leo_temporal_2010, coen_modeling_2013,
leo_dynamics_2013, coen_universal_2013, herr_temporal_2014, parra-rivas_dynamics_2014, godey_stability_2014}. These
studies have benefitted from the fact that KFCs can be modeled using a simple mean-field equation, the Lugiato-Lefever 
equation (LLE) \cite{coen_modeling_2013, chembo_spatiotemporal_2013}.

The bulk of KFC studies so far deals with microresonators exhibiting \emph{anomalous} second-order group velocity
dispersion (GVD) at the pump wavelength. However, due to the difficulty in obtaining anomalous GVD in some spectral
ranges, generation of KFCs from normal GVD microresonators is now also being sought and has recently been
achieved experimentally by several groups \cite{liang_generation_2014, huang_mode-locked_2015, xue_mode-locked_2015}.
In \cite{xue_mode-locked_2015}, a full time-domain characterization is reported: the field is found to consist
of square dark pulses of different widths --- low intensity dips embedded in a high intensity homogeneous background
--- with a complex temporal structure. These observations match several previous numerical predictions
\cite{matsko_normal_2012, coillet_azimuthal_2013, liang_generation_2014, godey_stability_2014,
lobanov_frequency_2015} and are in stark contrast with the isolated ultrashort bright localized structures observed
with anomalous GVD \cite{leo_temporal_2010, herr_temporal_2014}.

There has been some speculation as to the physical origin of the temporal structures observed in normal GVD KFCs,
which have been called platicons, dark pulse KFCs, or dark CSs \cite{lobanov_frequency_2015, xue_mode-locked_2015}.
To clarify this issue, we present here a detailed bifurcation analysis of dark structures in the LLE with normal GVD,
and predict their region of existence and stability. In particular, we clearly show that they are intimately related
to so-called switching waves (SWs) --- traveling front solutions of the LLE that connect the upper and lower
homogenous state solutions. These SWs were studied theoretically in the 80s \cite{rozanov_transverse_1982} and
their temporal dynamics were observed experimentally in fiber resonators in the 90s \cite{coen_convection_1999}. The
role of fronts in normal GVD KFCs is briefly suggested in \cite{lobanov_frequency_2015, xue_mode-locked_2015}, but in
fact it is so central to the stability of the dark structures that the term platicon does not really capture their
essence. In this Article, we therefore prefer to employ the terminology of dark pulse KFCs.

We start our analysis from the LLE, which was originally introduced in optics in the context of transverse resonators
\cite{lugiato_spatial_1987}. Using the normalization of \cite{leo_temporal_2010}, the LLE reads
\begin{equation}
  \label{eq.1}
  \partial_{t}u=-(1+i\theta)u+i|u|^{2}u+u_{0}-i\partial_{\tau}^{2}u\,,
\end{equation}
where $t$ is the slow time describing the evolution of the intracavity field $u(t,\tau)$ on the time scale of the cavity
photon lifetime, while $\tau$ is the fast time that describes the temporal structure of that field on the time scale of 
a resonator roundtrip $L$. The first term on the right-hand side describes cavity losses (the system is dissipative by 
nature); $\theta$ measures the cavity detuning between the frequency of the driving field and the nearest cavity 
resonance; the cubic term represents the (self-focusing) Kerr nonlinearity; $u_0$ is the amplitude of the homogeneous 
(cw) driving field or pump; and the fast-time derivative models GVD, here assumed normal at the pump frequency and 
limited to second order. In this form, the LLE has only two control parameters: $\theta$ and $u_0$.

\begin{figure}
\begin{center}
\includegraphics[width=\columnwidth]{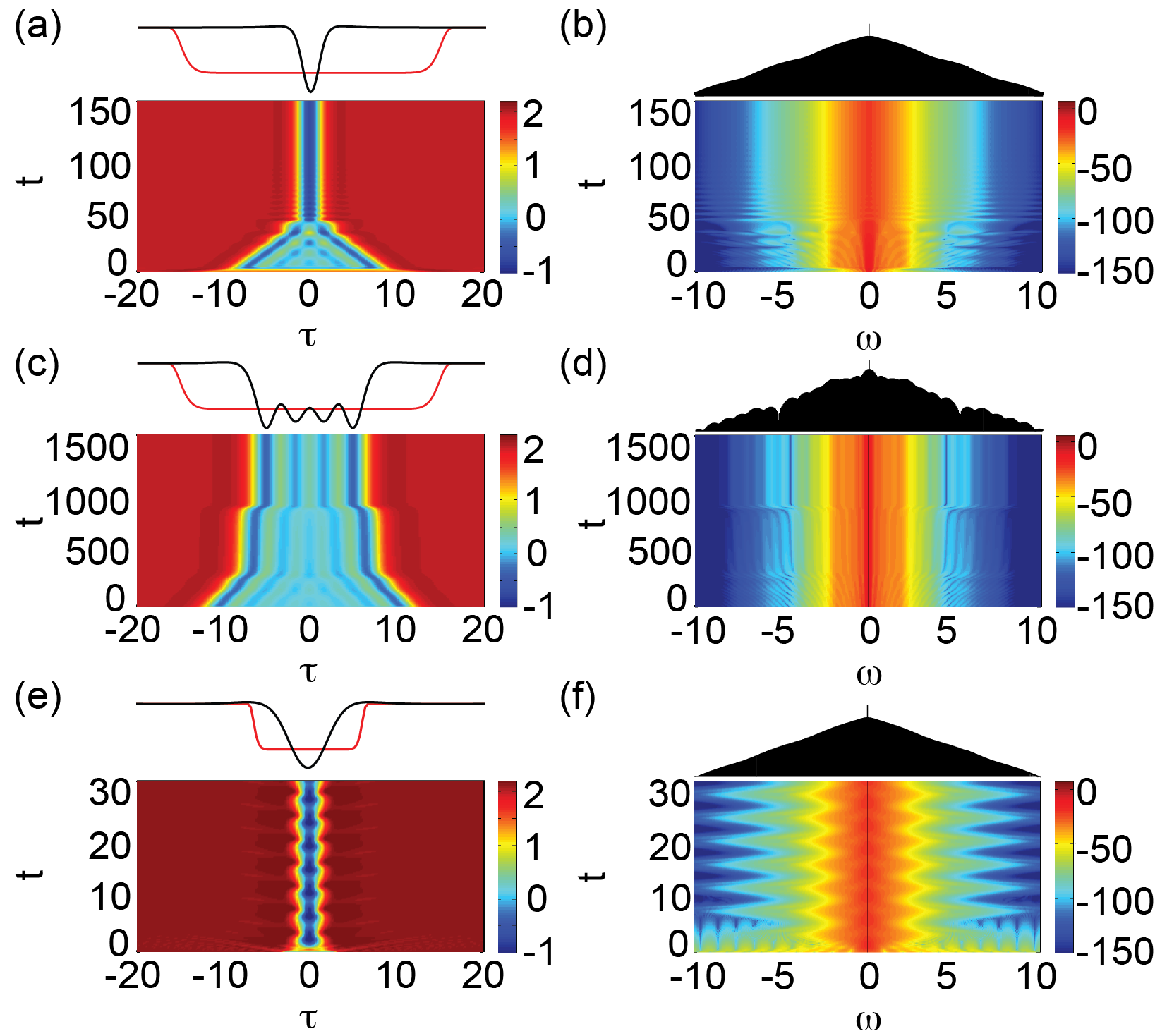}
\caption{Evolution of the temporal structure [left; $\mathrm{Re}(u)$] and corresponding spectral intensity (right; in dB) of an
          initial broad dark pulse over successive roundtrips represented as color density plots for (a,b) $(\theta, u_{0}) =
          (4, 2.25) $, (c,d) $(\theta, u_{0}) = (4, 2.18) $, (e,f) $(\theta, u_{0}) = (5, 2.6)$. The red (black) curves above
          the plots correspond to the initial (final) profiles.}
\label{time_evol_darkCS}
\end{center}
\end{figure}

To illustrate our discussion, we first present in Fig.~\ref{time_evol_darkCS} typical dark pulse KFC dynamics both in
the temporal (left) and spectral (right) domains. In each case, the initial condition (red curve atop each left plot)
is a broad dip of depth~2 in the upper homogeneous solution. In (a),(b) and (c),(d) we use the same detuning $\theta
= 4$ and initial condition but slightly different driving amplitude $u_0$ (as listed in the caption). In both cases
the initial broad dip shrinks in time, eventually coming to a halt and forming a stable dark pulse KFC. The final
dark temporal structures (black curve atop each plot) are markedly different, however, one being broader and far more
complex than the other. Fig.~\ref{time_evol_darkCS}(c) also reveals the intermediate existence of an even broader
structure. Next, Figs.~\ref{time_evol_darkCS}(e),(f) illustrate a simulation with an increased detuning, $\theta =
5$. The system again evolves to a dark pulse KFC, but it is no longer stable: it oscillates (breathes) in time. From
these simulations, several conclusions can be drawn about dark pulse KFCs (Refs.
\cite{godey_stability_2014, lobanov_frequency_2015, xue_mode-locked_2015} also mention these features): (i) they exist 
in the LLE with normal GVD and form as a dip in a background of higher intensity; (ii) they can have very different 
widths; and (iii) they can undergo dynamical instabilities such as breathing. In what follows, it is our goal to 
describe and explain all these different features and provide a map showing where the different solutions exist in the 
parameter space $(\theta, u_0)$.

\begin{figure}
\begin{center}
\includegraphics[width=\columnwidth]{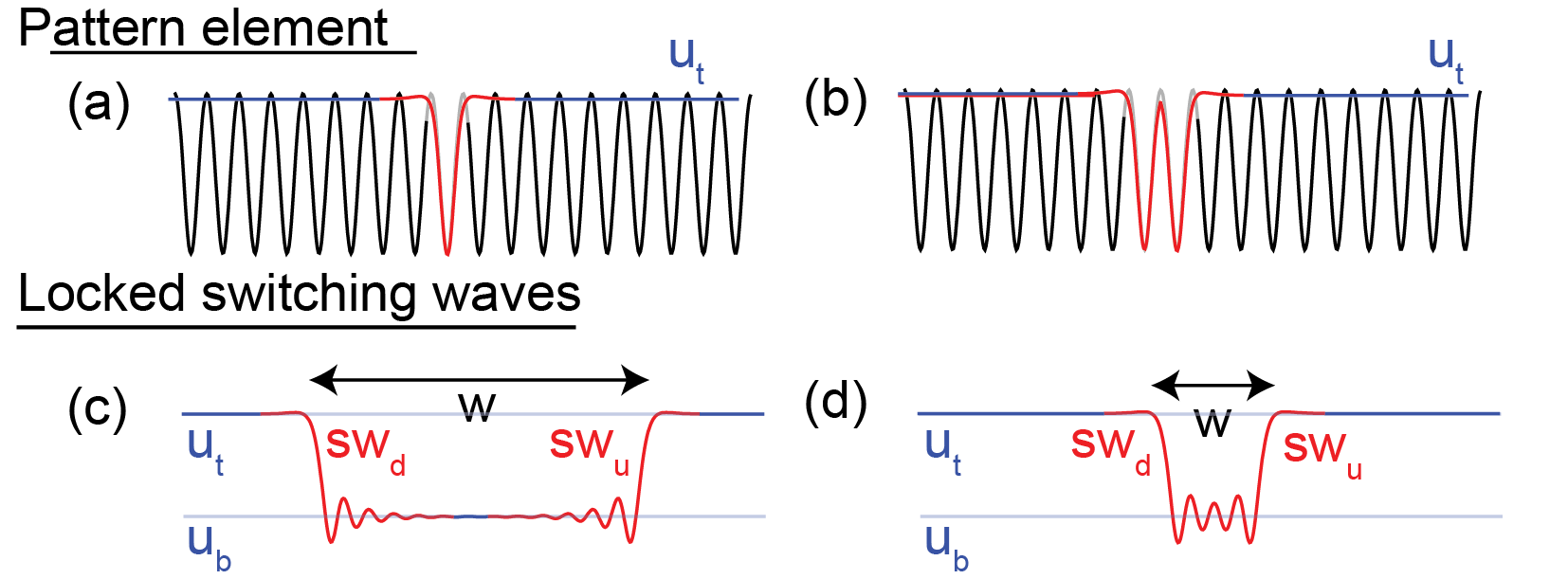}
\caption{Sketch of two different scenarios for dark pulse KFCs (in the time domain). (a),(b) \textit{Pattern element}:
    a connection between the high intensity homogeneous solution $u_\mathrm{t}$ and the periodic pattern.
    (c),(d) \textit{Locked SWs}: the high ($u_\mathrm{t}$) and low ($u_\mathrm{b}$) intensity homogeneous solutions
    form a stable connection.}
\label{qualitative_explanation_darksoliton}
\end{center}
\end{figure}

We start by providing a clear definition of a dark CS or dark pulse KFC and establishing the basic requirements for
their formation. Figs.~\ref{qualitative_explanation_darksoliton}(a),(b) sketch in the time domain the first scenario
(in red): a stable homogeneous steady state (HSS) solution of high intensity $u_\mathrm{t}$ (in blue) and a stable
periodic pattern generated at MI (in black) coexist and form a connection. As indicated, this connection can include one
oscillation of the pattern (a), two oscillations (b), or more. We refer to this type of dark pulse KFC as a
\textit{pattern element}. However, this type of connection does not appear to exist in Eq.~\ref{eq.1}, despite the coexistence
between a stable HSS and a periodic pattern for $\theta > 2$ \cite{haelterman_additive-modulation-instability_1992}.
In contrast, such localized solutions are present in LLE with anomalous GVD (albeit involving the \emph{low} intensity 
HSS $u_b$ rather than the high intensity HSS $u_t$) and explain the presence of the typical bright temporal CSs that 
are observed in this case \cite{leo_temporal_2010, leo_dynamics_2013, parra-rivas_dynamics_2014, godey_stability_2014}. 
In the presence of higher order GVD, the LLE with anomalous GVD can also admit a stable high intensity HSS and a periodic 
pattern, allowing dark CSs to exist as pattern elements \cite{tlidi_high-order_2010}.

Figs.~\ref{qualitative_explanation_darksoliton}(c),(d) demonstrate a second scenario for a dark pulse KFC. Whenever
two HSSs coexist, which occurs in Eq.~(\ref{eq.1}) for $\theta > \sqrt{3}$, and are both stable, which occurs with
normal GVD \cite{haelterman_additive-modulation-instability_1992}, the high ($u_\mathrm{t}$) and low ($u_\mathrm{b}$)
intensity HSSs (in blue) can connect to one another through what are called \textit{switching waves} (SWs) or fronts
\cite{rozanov_transverse_1982, pomeau_front_1986, coen_convection_1999}. When these SWs ($sw_\mathrm{d}$ and
$sw_\mathrm{u}$ in red) interlock stably, a dark temporal structure of width $w$ is formed. This is the type of dark
pulse KFC observed in microresonators with normal GVD.

\begin{figure}
\begin{center}
\includegraphics[width=\columnwidth]{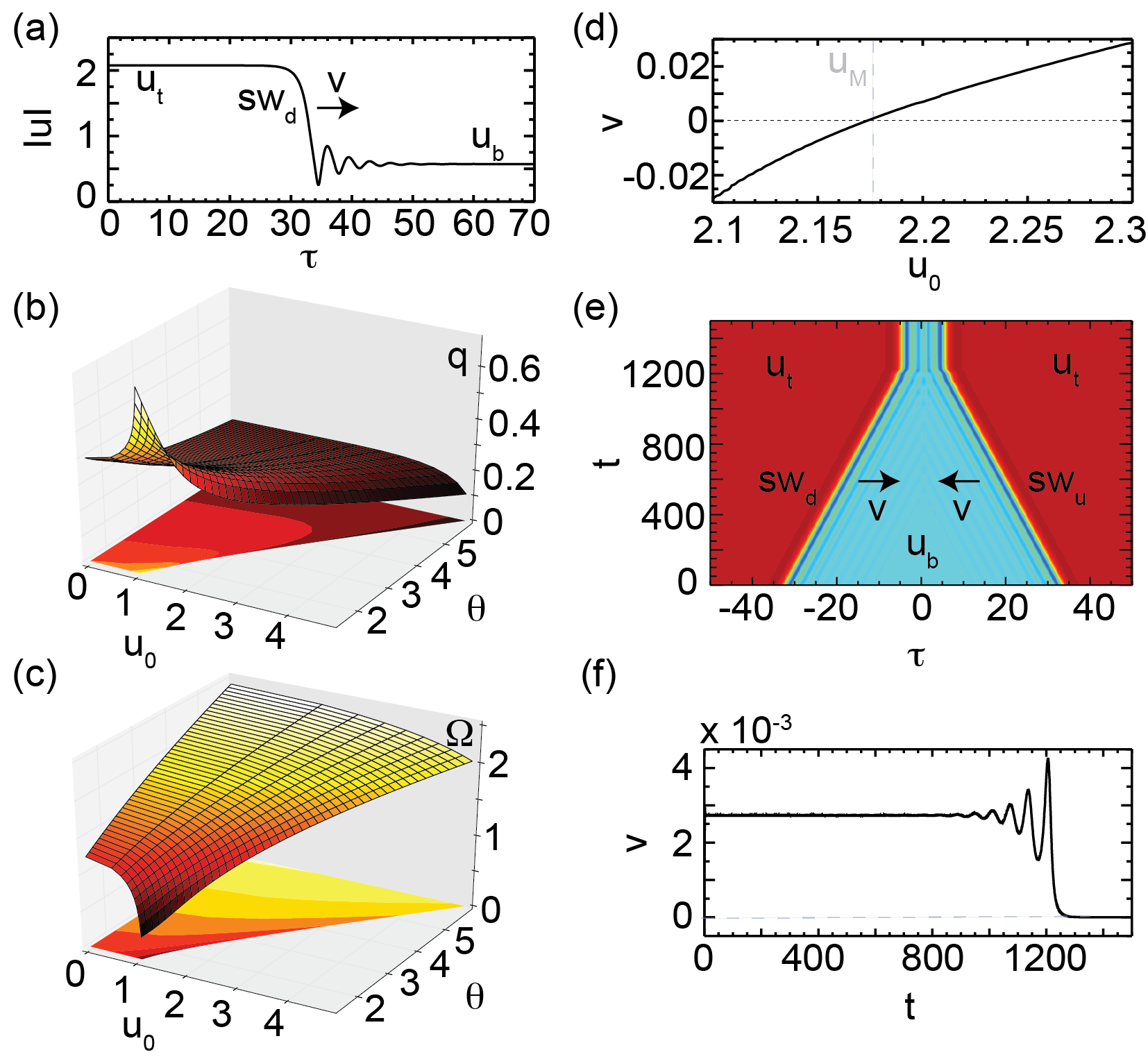}
\caption{(a) Temporal profile of $sw_\mathrm{d}$ for $(\theta, u_0)$ = $(4,2.175)$. (b,c) Damping rate $q$ and
          frequency $\Omega$ of the oscillatory tail near $u_\mathrm{b}$ as a function of $(\theta, u_0)$. (d) Velocity of
          $sw_\mathrm{d}$ versus driving amplitude $u_0$ for $\theta = 4$. (e) Temporal evolution of two SWs, $sw_\mathrm{d}$
          and $sw_\mathrm{u}$, approaching each other [density plot of $\mathrm{Re}(u)$] and (f) the corresponding velocity of
          $sw_\mathrm{d}$ over time.}
\label{switching_waves}
\end{center}
\end{figure}

To understand how such combinations of SWs arise and are stable, let us discuss the properties of individual SWs in
more detail (see also \cite{rozanov_transverse_1982, pomeau_front_1986, coen_convection_1999}).
Fig.~\ref{switching_waves}(a) shows the temporal profile $|u(\tau)|$ of a typical SW $sw_\mathrm{d}$ for $(\theta,
u_0)$ = $(4,2.175)$. The profile is asymmetric: $sw_\mathrm{d}$ leaves the top HSS $u_\mathrm{t}$ monotonically, but
approaches the bottom HSS $u_\mathrm{b}$ in an oscillatory way. Close to $u_\mathrm{b}$, this oscillatory approach
can be approximated linearly in the form $sw_\mathrm{d}(\tau) - u_\mathrm{b} \propto e^{\lambda \tau}$. Here $\lambda
= q + i \Omega$ is a complex eigenvalue, with $q$ the damping rate of the oscillations and $\Omega$ their frequency.
Figs.~\ref{switching_waves}(b),(c) illustrate how the eigenvalues, and thus the profile of the SW, change with the
parameters $(\theta, u_0)$. See \cite{colet_formation_2014, gelens_formation_2014} for more details on such
eigenvalue analysis. The SW generally moves (in the reference frame of the driving field) with
a velocity $v$ that depends on the driving amplitude $u_0$ [see Fig.~\ref{switching_waves}(d)]; $v = 0$ for only one
particular value of the driving amplitude, $u_0 = u_\mathrm{M}$, known as the \textit{Maxwell point} (for $\theta
=4$, $u_\mathrm{M} \approx 2.175$). The reason the SW moves can be intuitively understood as follows.
Above the Maxwell point, the system is comparatively closer to up-switching, thus $u_\mathrm{b}$ is less favorable,
and $u_\mathrm{t}$ invades $u_\mathrm{b}$. The opposite occurs below the Maxwell point. Given these properties, it
may seem that a stable dark pulse KFC made up of two SWs can only exist at the Maxwell point where the domains do not
invade each other. However, away from $u_\mathrm{M}$, as the SWs set into motion, their oscillatory tails may become
interlocked, as in other types of soliton bound states 
\cite{pomeau_front_1986,malomed_bound_1993,schapers_interaction_2000,barashenkov_interactions_2007,knobloch_spatially_2012,knobloch_spatially_2015}. This is
illustrated in Figs.~\ref{switching_waves}(e),(f): two SWs initially move towards each other at fixed mean velocity,
invading the $u_\mathrm{b}$ domain, but the overlapping oscillatory tails provide a counteracting
force that modulates their velocity, eventually bringing them to a halt and creating a stable dark pulse KFC. Dark
pulse KFCs can thus exist provided the velocity of the SWs is small enough to be balanced by the oscillations in the
SW tails. Note that the monotonic behavior of the SWs near $u_\mathrm{t}$ prevents ``bright'' combinations.

\begin{figure}
\begin{center}
\includegraphics[width=\columnwidth]{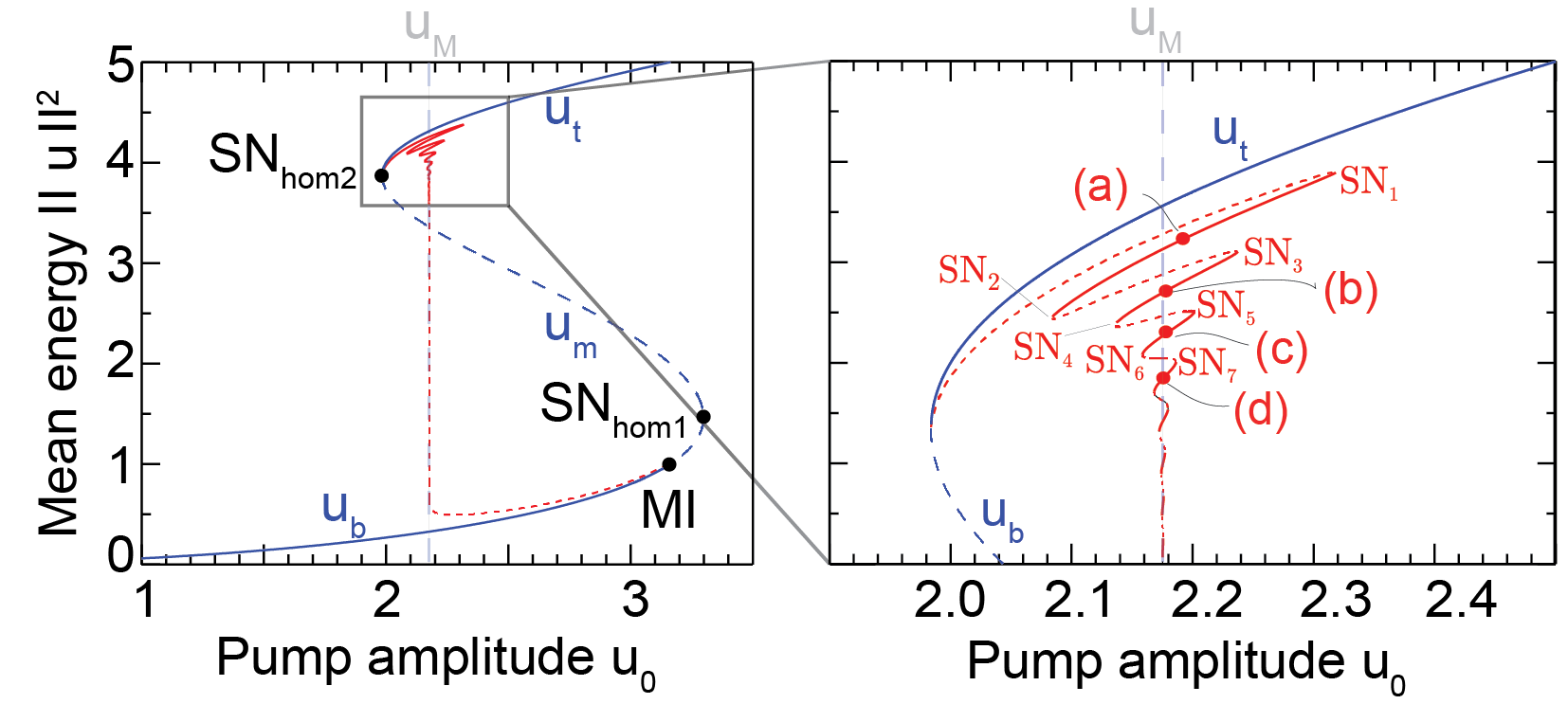}
\caption{Bifurcation diagram of dark pulse KFCs (red) for $\theta=4$, with a zoom (right) of the bifurcation branches close
          to the top HSS $u_t$. Stable (unstable) structures are indicated using solid (dashed) lines. Blue lines
          correspond to HSSs.}
\label{bifurcation_structure_det_4}
\end{center}
\end{figure}

\begin{figure}
\begin{center}
\includegraphics[width=\columnwidth]{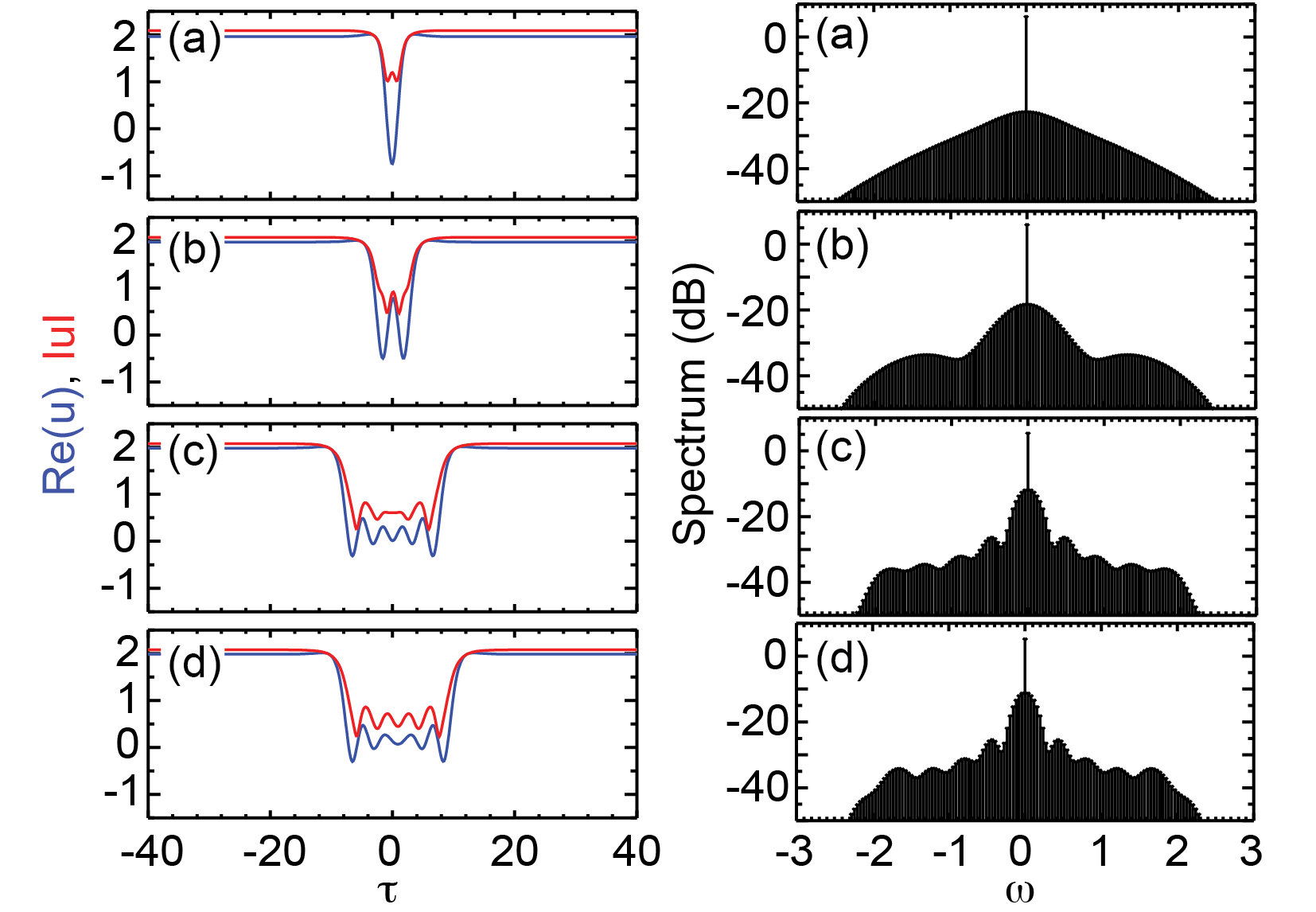}
\caption{Temporal profiles [left; $\mathrm{Re}(u)$ in blue, $|u|$ in red] and spectral intensities (right, in dB) of
          dark pulse KFCs corresponding to the locations (a)--(d) in Fig. \ref{bifurcation_structure_det_4}.}
\label{profiles_det_4}
\end{center}
\end{figure}

Fig.~\ref{bifurcation_structure_det_4} shows the bifurcation diagram of the dark pulse KFCs (in red) obtained by
numerical continuation. The figure shows, in blue, the top ($u_\mathrm{t}$), middle ($u_\mathrm{m}$), and bottom 
($u_\mathrm{b}$) HSSs: these are solutions of $I_\mathrm{t,m,b}^3-2\theta I_\mathrm{t,m,b}^2+(1+\theta^2)I_\mathrm{t,m,b}=u_0^2$, 
where $I_\mathrm{t,m,b}\equiv|u_\mathrm{t,m,b}|^2$. These three solutions only exist for $\theta>\sqrt{3}$, and the 
transitions between them occur via two saddle-node bifurcations SN$_\mathrm{hom,1}$ and SN$_\mathrm{hom,2}$. The 
diagram shows that unstable dark KFCs originate from SN$_\mathrm{hom,2}$ and that as $u_0$ increases their mean energy
$||u||^2\equiv L^{-1}\int_0^{L}|u|^2\,d\tau$ grows until SN$_1$, a saddle-node bifurcation at which they acquire 
stability. The temporal profile on the resulting segment of stable dark solitons is shown in Fig.~\ref{profiles_det_4}(a). 
These KFCs lose stability at SN$_2$ but start to develop an additional oscillation in the center. Solutions of this 
type become stable at SN$_3$; an example of the resulting stable double-oscillation dark pulse KFC can be found in 
Fig.~\ref{profiles_det_4}(b). This process repeats in such a way that between successive saddle-nodes on the left or 
right a new oscillation period is inserted in the center of the dark structure, resulting in a structure that is 
temporally broader but less energetic [see Figs.~\ref{profiles_det_4}(c),(d)].

An additional feature of this snaking branch of solutions is that after each successive turn the range of driving
amplitudes over which the branch of solutions exists becomes narrower. In other words, the temporally broader the
dark pulse KFC is (or equivalently, the more oscillations it has), the narrower its range of existence. This type of
organization is thus appropriately referred to as \textit{collapsed snaking}. This collapse originates in the
exponential damping of the oscillations of the SWs: As the distance between $sw_\mathrm{u}$ and $sw_\mathrm{d}$
increases, the oscillations rapidly become too weak to be able to compensate the inherent velocity of the SWs.
Therefore, very temporally broad dark pulse KFCs can only exist at the Maxwell point $u_\mathrm{M}$.

\begin{figure}
\begin{center}
\includegraphics[width=\columnwidth]{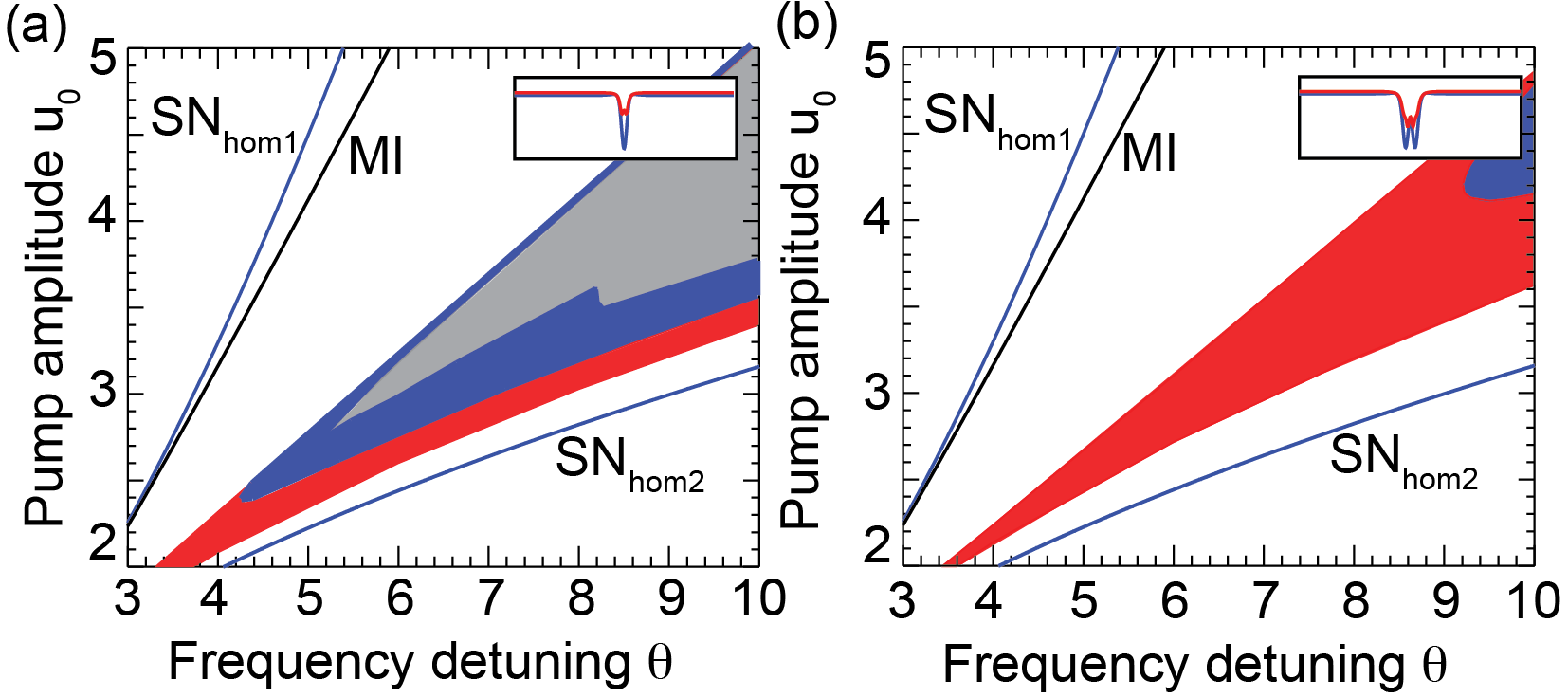}
\caption{Regions of existence in the $(\theta,u_0)$ parameter space of (a) the single- and (b) the double-oscillation
          dark KFC. The red (blue) region corresponds to stable stationary (oscillatory) dark KFCs, while
          within the gray region dark KFCs are unstable and relax to the HSS after a chaotic transient.}
\label{dynamics}
\end{center}
\end{figure}

The organization of dark pulse KFCs in a collapsed snaking structure strongly suggests that the narrowest dark pulse
KFCs should be the most frequent occurrence in experiments. Although this is true for a wide range of parameter
values, this statement requires modification as the detuning $\theta$ increases, because dark pulse KFCs can then
undergo a wide range of instabilities similar to those of bright CSs in the case of anomalous GVD
\cite{leo_dynamics_2013, parra-rivas_dynamics_2014}. In Figs.~\ref{time_evol_darkCS}(e),(f), we already saw one
example of an unstable (breathing) dark pulse KFC. We have explored these instabilities further and Fig.~\ref{dynamics} 
shows stability maps of (a) single-oscillation and (b) double-oscillation dark pulse KFCs. Examples of their temporal 
profiles, corresponding to Figs.~\ref{profiles_det_4}(a),(b), are shown in insets. As a reference, we plot the 
$SN_\mathrm{hom,1}$, $SN_\mathrm{hom,2}$, and $MI$ bifurcation lines because dark pulse KFCs can only exist between the 
$SN_\mathrm{hom,2}$ and $MI$ lines where $u_\mathrm{t}$ and $u_\mathrm{b}$ are both stable (Fig.~\ref{bifurcation_structure_det_4}).
The red regions in Fig.~\ref{dynamics} show the parameter values for which single-oscillation and double-oscillation 
dark pulse KFCs can be found and are stable. Although the region of existence of dark pulse KFCs shrinks as they become 
temporally broader (increasing the number of oscillations), the range of stable KFCs is broader for broader KFCs:
the stable red region is larger for the double-oscillation KFC than the single-oscillation KFC. In fact dark pulse 
KFCs with three or more oscillations are stable across their whole region of existence at least up to values of the 
detuning as high as~10 (not shown here).

As shown in Figs.~\ref{time_evol_darkCS}(e),(f), for $(\theta, u_{0}) = (5, 2.6)$ the single-oscillation KFC is no
longer stationary, but instead breathes in time with a well-defined period and amplitude. Such breathers are found
throughout the blue regions in Fig.~\ref{dynamics}. Within the blue region, but close to the gray region, period
doubling occurs and breathing with multiple periods can be observed, eventually becoming chaotic. In all of these
cases, the dark pulse KFC remains localized. In the gray region, however, dark pulse KFCs lose stability to
chaotic but transient oscillations, and eventually collapse to the HSS. More details about these dynamics and a more elaborate bifurcation analysis will be presented in \cite{rivas_dark_long_2016}.

In summary, we have provided a bifurcation analysis of dark pulse KFCs and mapped out their existence and stability
conditions, as well as some of their dynamical instabilities. The KFCs are shown to be organized in a collapsed snaking
structure resulting from the oscillatory tails of the SW that form them. Many observed properties of dark pulse KFCs, 
including coexistence of dark structures of different widths, can be similarly explained based on known features of 
SW dynamics. We note that MI, despite occurring in normal GVD resonators \cite{coen_competition_1999}, does not play
a crucial role in this description: the SWs connect the upper and lower HSSs and are not related to extended patterns. 
Some form of MI involving different microresonator mode families is, however, likely to be involved in explaining the 
soft excitation of dark pulse KFCs \cite{lobanov_frequency_2015,xue_mode-locked_2015}.

We acknowledge support from the Research Foundation--Flanders (FWO-Vlaanderen) (PPR), the KU Leuven Junior Mobility
Programme (LG), the Belgian Science Policy Office (BelSPO) under Grant IAP 7-35, the Research Council of the Vrije
Universiteit Brussel, the Spanish MINECO and FEDER under Grant Intense@Cosyp (FIS2012-30634) (DG), the National Science Foundation under grant DMS-1211953 (EK), and the Marsden Fund of the Royal Society of New Zealand (SC). We also thank F. Leo for valuable discussions.

\end{document}